\documentclass{raa}
\usepackage[utf8]{inputenc}
\usepackage[T1]{fontenc}
\usepackage{graphicx}
\usepackage{longtable}
\usepackage{wrapfig}
\usepackage{rotating}
\usepackage{amsmath}
\usepackage{amssymb}
\usepackage{capt-of}
\usepackage{hyperref}
\usepackage{graphicx,times}             
\usepackage{natbib}
\usepackage{float}
\usepackage{longtable}
\bibpunct{(}{)}{;}{a}{}{,}
\author{Yi Yang
\inst{1}
\and Zhenzhen Shao\
\inst{1,2,3}
\and Xiaofeng Wang
\inst{4}
\and Xiaochen Zheng
\inst{1}
}
\institute{Department of Scientific Research, Beijing Planetarium, Beijing Academy of Science and Technology, Beijing 100044, China; {\it mgcyung@gmail.com}\\
\and
Institute for Frontiers in Astronomy and Astrophysics, Beijing Normal University, Beijing 102206, China; {\it 201011161010@mail.bnu.edu.cn} \\
\and
School of Physics and Astronomy, Beijing Normal University, Beijing 100875, China; \\
\and
Physics Department, Tsinghua University, Beijing 100084, China;  {\it wang\_xf@mail.tsinghua.edu.cn}\\
\vs\no
{\small Received 20xx month day; accepted 20xx month day}}
\abstract{  This study introduces an automated approach for identifying the tip of the red giant branch (TRGB) in globular clusters, combining astronomical data with algorithmic methods. Using a dataset of 160 globular clusters and Python scripts, we matched stellar sources with Gaia data. Our script generates color-magnitude diagrams (CMDs), and uses the local outlier factor (LOF) algorithm to remove outliers. Applying a second-degree polynomial to fit red giant branch (RGB), we identify the TRGB as the star closest to the fitted curve's endpoint. By this methdod, we expanded TRGB samples in global clusters to 91 with newer observational data. Our results show a decreasing trend in I-band luminosity for metallicities greater than $-$1, consistent with previous studies.  The results show a robust trend fitting and the $\rm M_{I}$ of TRGB is about \(-\)4.02 with extremely low metallicity. Our approach enhances TRGB identification efficiency while providing valuable insights for developing automatic tools in astronomical data analysis.  }
\date{}
\title{Automated Identification of the Tip of the Red Giant Branch in Globular Clusters with Gaia Data}
\titlerunning{~}
\begin{document}

\maketitle
\section{Introduction}
\label{sec:org8db109d}

The tip of the red giant branch (TRGB) marks the maximum luminosity
attained by stars as they evolve along the red giant branch (RGB)
in the Hertzsprung-Russell diagram (HRD). Characterized by its
consistent I-band absolute magnitude, the TRGB serves as a valuable
standard candle for distance estimation in nearby galaxies
\citep{2019-Freedman-CarnegieChicago-A,2021-Hoyt-Carnegie-A,2001-Bellazzini-Step-A,2007-Rizzi-Tip-A,2000-Sakai-Hubble-A},
with growing applications in Hubble constant measurements
\citep{2021-Freedman-Measurements-A,2019-Freedman-CarnegieChicago-A,2019-Yuan-Consistent-A,2021-Anand-Distances-M,2021-Soltis-Parallax-A}.
Previous investigations have revealed metallicity dependence
of the TRGB's absolute magnitude. Studies by \cite{1997-Salaris-Tip-M}
and \cite{2017-Serenelli-Brightness-A} established I-band
luminosity-metallicity relationships through stellar evolution modeling,
while \cite{1990-DaCosta-Standard-A} and \cite{2000-Ferraro-New-A}
examined relationships between each of bolometric magnitudes, color indices,
bolometric corrections and metallicity. Building on this foundation,
\cite{2001-Bellazzini-Step-A} and \cite{2017-Jang-TIP-A} developed
empirical relations between I-band absolute magnitude and metallicity.

Current TRGB detection methodologies primarily employ edge detection
algorithms and parametric fitting approaches. The Sobel filter method
\citep{2017-McQuinn-DUSTiNGS-A,2017-Jang-TIP-A,2023-Scolnic-CATS-A} identifies luminosity
function discontinuities through kernel-based gradient estimation,
though its performance is sensitive to smoothing parameter
optimization. Alternative techniques fit broken power-law models to the
luminosity function \citep{2023-Li-Gaia-A,2024-Li-Tip-}, but these
assume a single discontinuity in the RGB structure.

Globular clusters, provide an ideal environment for TRGB
studies. Their color-magnitude diagrams (CMDs) typically encompass
complete stellar evolutionary sequences from main sequence through red
giant to horizontal branch stages. The near-uniform age and
metallicity of cluster members - primary factors influencing TRGB
luminosity - offer particularly favorable conditions for
analysis. However, due to the limited number of RGB stars in
individual clusters, conventional edge detection and parametric
fitting approaches prove less effective for TRGB identification
\citep{2004-Bellazzini-Calibration-A}.

TRGB corresponds to the theoretical evolutionary track of the RGB and
exhibits the highest luminosity in globular clusters. Nevertheless,
the conversion of luminosity to magnitude is influenced by effective
temperature, surface gravity, metallicity, and bolometric correction,
which means that the highest point of luminosity may not correspond to
the brightest magnitude on a specific band in the
CMD. \cite{2025-Shao-Dependence-A} referred that TRGB corresponds to
the reddest star rather than brightest one in globular clusters.

Building on previous work, \cite{2025-Shao-Dependence-A} demonstrated
that M\textsubscript{I} remains nearly constant for [Fe/H] < \(-\)1.2 but decreases
with higher metallicity, using TRGB stars identified through selection
of the reddest star in G\textsubscript{BP}\(-\)G\textsubscript{RP} versus G\textsubscript{RP} diagrams from
33 Galactic globular clusters. However, this metallicity relation
remains constrained by limited sample sizes in metal-rich regimes
([Fe/H] > \(-\)1.2).

Extending the reddest RGB selection method of
\cite{2025-Shao-Dependence-A}, we implement an automated pipeline
incorporating algorithmic filtering, outlier removal, and curve
fitting for TRGB identification. This automated approach enables
efficient processing of extensive datasets while maintaining selection
consistency, reducing manual intervention, and enhancing result
reproducibility. By automating the identification of TRGB stars in
globular clusters, the method can be used for TRGB studies in large
scale data in CSST and JWST \citep{2024-Anand-TRGBSBF-}.
\section{Data}
\label{sec:orga36749f}

The data employed in this study derive from the globular cluster
star catalog compiled by \cite{2021-Vasiliev-Gaia-M}. This
catalog, combined with contemporary algorithmic techniques,
facilitates precise TRGB star determination within globular clusters.
The primary data source encompasses 160 globular clusters, as detailed
in \cite{2021-Vasiliev-Gaia-M}. Rigorous
quality control measures are employed to select stars with reliable astrometric
parameters. These refined subsets were then used to ascertain cluster
properties and individual stellar membership probabilities through
a mixed modeling approach extensively described in Section 2 of their
work.

To achieve comprehensive RGB source coverage, we initially set the
membership probability threshold at 90\%. Subsequently, to minimize
non-member star inclusion, we increased the threshold to 99\%. We
conducted a comparison between 90\% and 99\% membership probability
thresholds, finding that most stars removed during filtering resided
in the main sequence as the threshold increased from 90\% to 99\%, with
statistical impact on RGB star selection being negligible for over 95\%
of clusters. However, five clusters showed partial RGB star exclusion
under the stricter 99\% criterion.  Therefore, we selected 99\% as the
membership probability threshold, excluding these clusters from our
final sample to maintain high selection standards. This stringent
criterion aims to achieve datasets containing only confirmed cluster
members \citep{2021-Vasiliev-Gaia-M}.

To enhance photometric information, we cross-matched our data with the
Gaia DR3 catalog \citep{2023-GaiaCollaboration-Gaia-A}. This process
provided BP (G\textsubscript{BP}) and RP (G\textsubscript{RP}) magnitudes for sample stars,
which offer crucial color information for analysis.

We further augmented our dataset by cross-matching with
\cite{2019-Stetson-Homogeneous-MNRAS} and the SIMBAD astronomical
database \citep{2000-Wenger-SIMBAD-AASS}. While most data originated
from \cite{2019-Stetson-Homogeneous-MNRAS}, we queried SIMBAD for
missing sources.  This cross-matching process enabled retrieval of
available V- and I-band magnitudes. Additionally, we supplemented
these data through consultation of astronomical literature. When
available, these magnitudes are crucial for accurate CMD construction
and analysis.
\section{Method and Data Processing}
\label{sec:org3eae817}

\begin{figure}[!t]
\centering
\includegraphics[width=.9\linewidth]{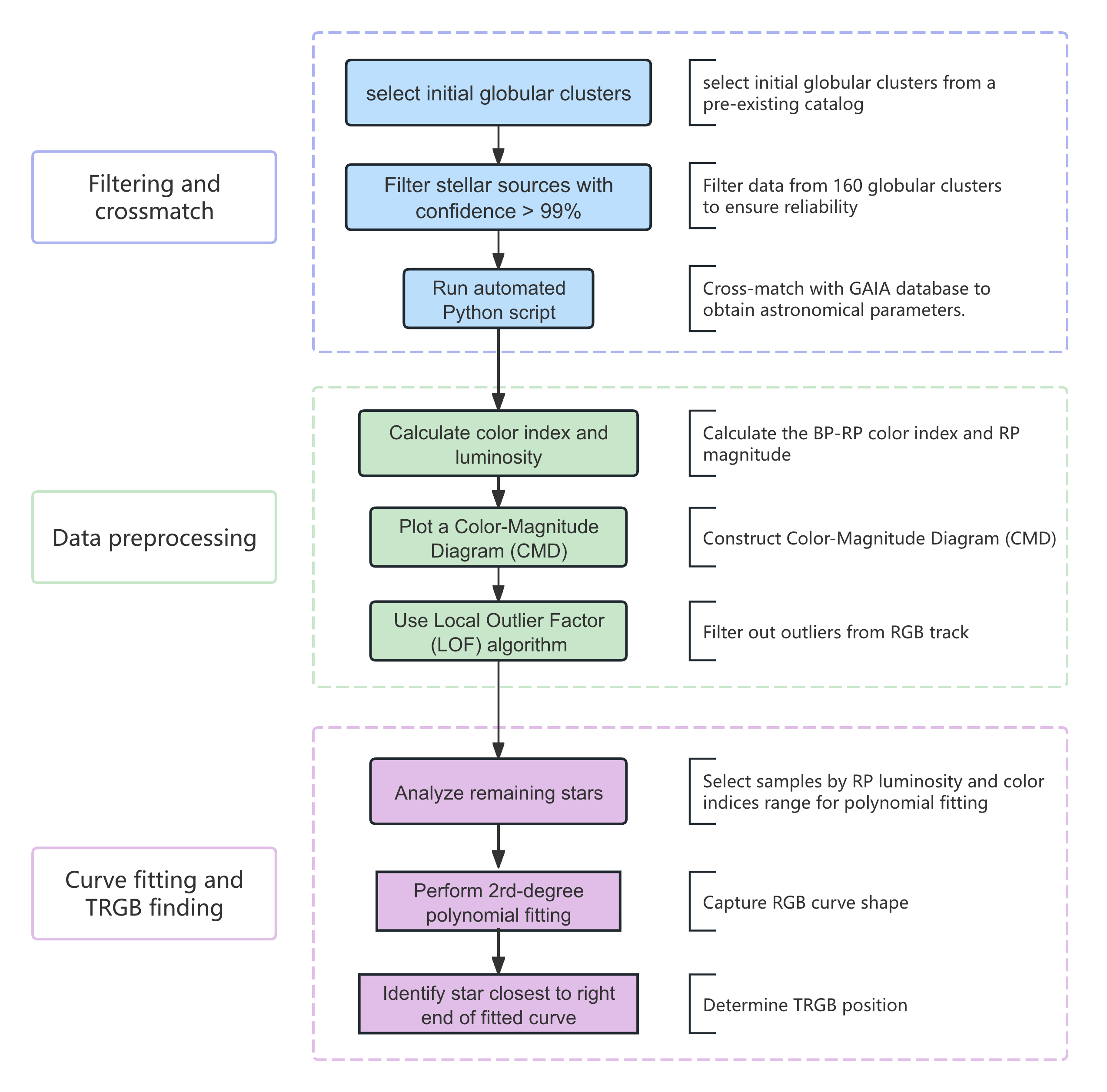}
\caption{\label{fig:architecture-TRGB}
Diagram of the TRGB seeking procedure, which includes data filtering and cross-referencing, data preprocessing, and curve fitting and TRGB finding.}
\end{figure}

As shown in Figure \ref{fig:architecture-TRGB}, our method primarily consists
of three components: data filtering and cross-referencing, data
preprocessing, and curve fitting and TRGB finding.

\subsection{Filtering and crossmatching}
\label{sec:org6407d80}

As shown in Figure \ref{fig:architecture-TRGB}, the data filtering and
cross-matching process comprises three key steps: cluster selection,
stellar member selection, and Gaia database cross-matching.

We begin with an initial sample of 160 globular clusters from an
existing catalog. Initial number of stars ranged from several dozen to
several hundred thousand stars per cluster, depending on richness and
observational coverage. We then filter stellar sources within these
clusters by selecting those with membership probabilities exceeding
99\%, which minimized field star contamination while maintaining
sufficient sample sizes (ranging from dozens to tens of thousands of
stars per cluster) for reliable analysis. As shown in Figure
\ref{fig:result-TRGB}, most stars (grey) removed during filtering resided in
the main sequence as the threshold increased from 90\% to 99\%. Although
globular clusters exhibit multiple stellar populations primarily
affecting \(\alpha\)-element abundances (C, N, O, etc.)
\citep{2023-Leitinger-Widefield-M}, these variations negligibly impact
{[}Fe/H] measurements, making metallicity determinations robust against
population effects in our source selection.

Following this filtering step, we processed the data and performed a
cross-match with the Gaia DR3 catalog using an automated Python script
which utilizes astropy \citep{2022-TheAstropyCollaboration-Astropy-A}
and astroquery \citep{2019-Ginsburg-Astroquery-A}. This provides
detailed astrometric parameters for each source, including
high-precision positional, G magnitude, BP and RP magnitudes,
parallax, and proper motions from Gaia
\citep{2016-GaiaCollaboration-Gaia-A,2023-GaiaCollaboration-Gaia-A},
which are essential for further analysis in TRGB identification.
\subsection{Data preprocessing}
\label{sec:orgb9f65ee}

As shown in Figure \ref{fig:architecture-TRGB}, the data preprocessing stage
involves three main operations: BP-RP color index calculation, CMD
construction, and local outlier factor (LOF) filtering
\citep{2000-Breunig-LOF-P2ASICMD}.

Using Gaia data, we constructed CMDs for each cluster with BP-RP color
indices and RP magnitudes which reveal stellar evolutionary
stages. The RGB appears as a distinct sequence in these diagrams, with
the TRGB positioned at its peak. As shown in Figure \ref{fig:result-TRGB},
these CMDs clearly delineate the red giant branch while visualizing
stellar distributions, with typical ranges of \(-\)1.0 < BP\(-\)RP <
5.0 and 10 < RP < 22 respectively.

After constructing the CMDs, we implemented a further culling step
based on visual inspection of preliminary CMDs. We excluded globular
clusters that either have a too small number of stars (such as NGC
6528), have a too small number of stars on RGB (such as NGC 6535), or
have a dispersed RGB (such as NGC 6287) indicative of compromised data
quality. This step ensured that subsequent analysis focused on
clusters with sufficient data of adequate quality to allow for robust
TRGB identification.

To improve TRGB identification accuracy, we apply the LOF algorithm to
remove outliers from the RGB sequence. This unsupervised anomaly
detection method identifies points that deviate significantly from
their local neighborhood structure. Using parameters n\_neighbors=20
and contamination=0.005, the filter effectively eliminates outliers
which may have measurement errors while preserving the intrinsic RGB
samples. The LOF algorithm removed 0-2.1\% of outliers (average 0.4\%)
that deviated from CMD neighborhood structures, primarily eliminating
potential measurement errors or non-member stars while preserving RGB
morphology. As shown in Figure \ref{fig:result-TRGB}, these filtered
outliers (green) typically occupied peripheral CMD regions, enhancing
RGB clarity in the remaining sample (blue).
\subsection{Curve fitting and TRGB identification}
\label{sec:org939592f}

As depicted in Figure \ref{fig:architecture-TRGB}, the curve fitting and
TRGB identification process involves three sequential steps:
luminosity range–based sample selection, polynomial fitting, and TRGB
determination.

After cleaning the data, we performed polynomial fit to stellar
sources in each cluster. We first select stars with RP magnitudes no
less than 3.0 mag of the brightest star on the red giant branch and
with BP-RP color indices greater than 1.2 for polynomial fitting. This
selection focuses on the luminous and red region of the RGB where the
TRGB is expected to reside. A second-degree polynomial is then applied
to fit on these selected samples using a HuberRegressor
\citep{2007-Owen-Robust-CM}, with BP\(-\)RP color index as the
independent variable and RP magnitude as the dependent variable.  This
robust regression approach effectively balances curve representation
and overfitting prevention, providing a reliable model of the RGB
morphology.

\begin{figure}[!t]
\centering
\includegraphics[width=.9\linewidth]{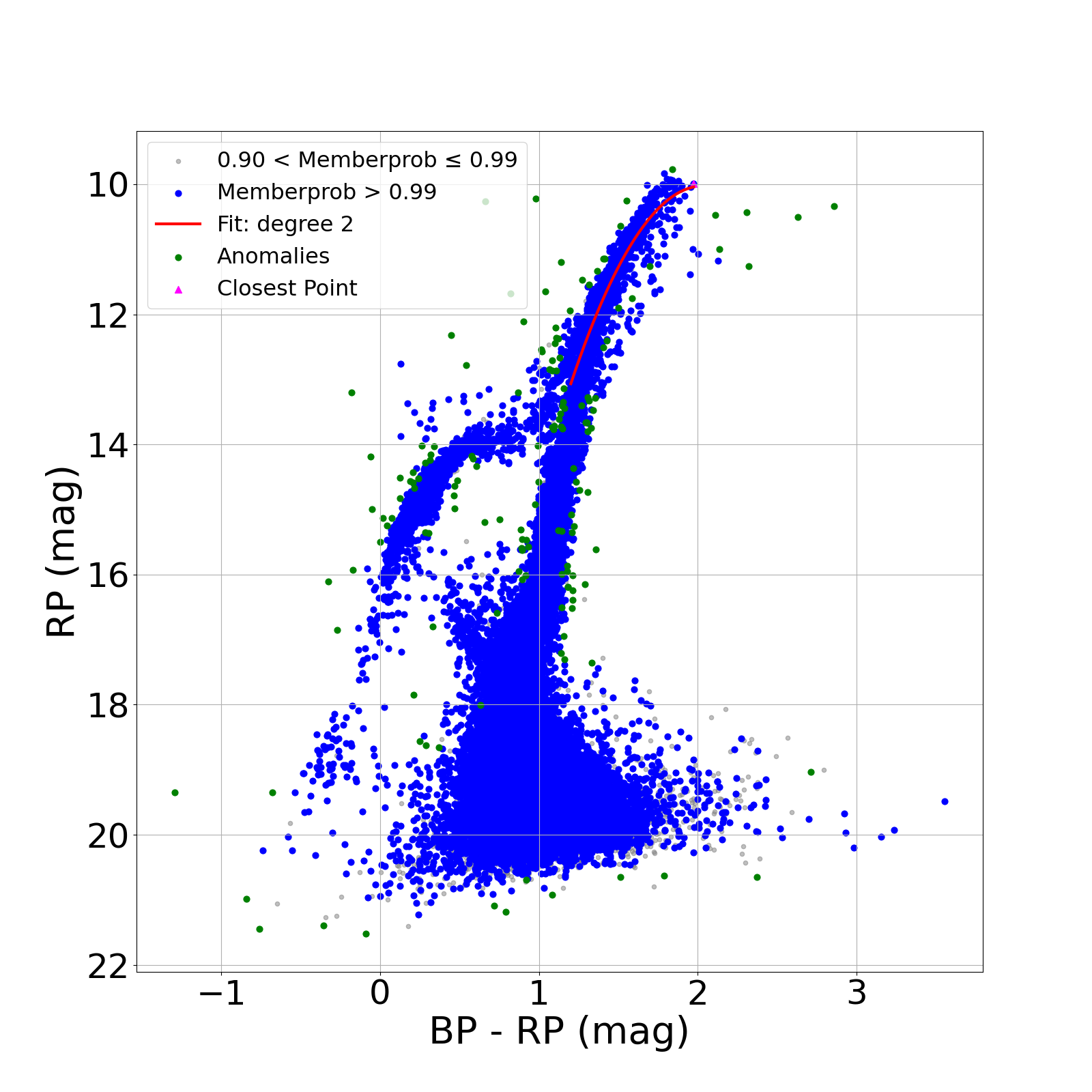}
\caption{\label{fig:result-TRGB}
Color-magnitude diagram for NGC 5139, showing sources with memberprob greater than 99\% (blue dots), sources with memberprob in range from 90\% to 99\%, polynomial fit (red curve), outliers removed by LOF (green dots), and the identified TRGB candidate (magenta triangle).}
\end{figure}

The resulting polynomial curve effectively captured the overall trend
of the red giant branch, which is a red curve in Figure
\ref{fig:result-TRGB}. The TRGB candidate was identified as the star
'closest' to the right end of the fitted curve, corresponding to the
reddest terminus of the red giant branch, as shown in
\cite{2025-Shao-Dependence-A}. The closest means the weighted distance
of each sample to the reddest terminus of the fitted curve, assigning
a weight of 5.0 to the horizontal (BP\(-\)RP) axis and 1.0 to the
vertical (RP) axis. This objective procedure ensures consistent
identification of the RGB’s true apex while minimizing contamination
from anomalous sources. As shown in Figure \ref{fig:result-TRGB}, the TRGB
candidate was marked with a magenta triangle.

We examined our TRGB candidates and found that some were classified as
long-period variable stars (LPVs) on SIMBAD.
\cite{2024-Anderson-Smallamplitude-AJL} show that TRGB candidates
exhibit LPV characteristics, suggesting intrinsic brightness
fluctuations in TRGB stars. As the primary source of contamination
arises from asymptotic giant branch (AGB) stars. To mitigate this, we
applied the empirical criterion that I-band variability amplitudes of
AGB is larger than 0.14 mag \citep{2005-Lebzelter-Long-A}. We
identified only one source exceeding the threshold on Gaia RP-band
variability amplitudes\footnote{The I-band and RP wavelengths are close in proximity, allowing for the use of RP variability as an approximation for I-band variability.} by cross math with a catalog covers
magnitude variability of handred million sources
\citep{2023-MaizApellaniz-Stellar-A}. The identified source was
subsequently excluded from our final samples.

\section{Results and Discussion}
\label{sec:org473d809}

After applying filtering, cross-matching, data preprocessing, curve
fitting and TRGB identification, we ultimately obtained TRGBs for 94
globular clusters. We consider that they are the most possible TRGBs
in these global clusters. Below we present results and discussion.
\subsection{TRGB Results}
\label{sec:orgbaac7c6}

By utilizing Gaia parallax-based distance calibrations for Galactic
globular clusters from \cite{2021-Vasiliev-Gaia-M}, combined with the
color excess from \cite{2010-Harris-New-ae} and the extinction law of
\cite{2019-Wang-Optical-A}, the absolute magnitudes of the TRGB in the
V and I bands were determined. This enabled the derivation of the
relationship between the TRGB’s absolute magnitude and the metallicity
of globular clusters as cataloged in \cite{2010-Harris-New-ae}.

As shown in Figure \ref{fig:result-mag-meta}, 34 samples (green dots) show
consistency with the manual results \citep{2025-Shao-Dependence-A},
while 6 samples (red dots) display discrepancies. The TRGB points
identified by our method closely match the manual selection, achieving
an agreement of approximately 85\%, thereby demonstrating the accuracy
and reliability of our approach. Moreover, we extended 54 new samples
(black dots).

The astrometric and photometric properties of 94 TRGB stars are
summarized in Table \ref{tab:cluster_trgb}. Specifically, the table presents
the following information for each cluster: name, right ascension (RA)
and declination (Dec) for J2000.0, distance, E(B\(-\)V), [Fe/H], M\textsubscript{I}
(absolute magnitude of I band), and Err\textsubscript{I} (error of absolute
magnitude of I band, which is calculated by distance error,
photometric error, and extinction error).

\begin{longtable}{cccccccc}
\caption{\label{tab:cluster_trgb}
Catalogue of TRGBs in globular clusters.}
\\
\hline
Name & RA & DEC & Distance & E\textsubscript{(B-V)} & {[}Fe/H] & M\textsubscript{I} & Err\textsubscript{I}\\
 & (\(^{\circ}\)) & (\(^{\circ}\)) & (kpc) & (mag) &  & (mag) & (mag)\\
\hline
\endfirsthead
\multicolumn{8}{l}{Continued from previous page} \\

Name & RA & DEC & Distance & E\textsubscript{(B-V)} & {[}Fe/H] & M\textsubscript{I} & Err\textsubscript{I}\\
 & (\(^{\circ}\)) & (\(^{\circ}\)) & (kpc) & (mag) &  & (mag) & (mag) \\

\hline
\endhead
\hline\multicolumn{8}{r}{Continued on next page} \\
\endfoot
\endlastfoot
\hline
NGC 104 & 5.960 & \(-\)72.097 & 4.521 & 0.04 & \(-\)0.72 & \(-\)3.978 & 0.026\\
NGC 288 & 13.171 & \(-\)26.557 & 8.988 & 0.03 & \(-\)1.32 & \(-\)4.109 & 0.034\\
NGC 362 & 15.841 & \(-\)70.905 & 8.829 & 0.05 & \(-\)1.26 & \(-\)3.704 & 0.031\\
NGC 1261 & 48.010 & \(-\)55.180 & 16.4 & 0.01 & \(-\)1.27 & \(-\)4.002 & 0.025\\
NGC 1851 & 78.505 & \(-\)40.034 & 11.951 & 0.02 & \(-\)1.18 & \(-\)3.908 & 0.024\\
NGC 1904 & 81.038 & \(-\)24.571 & 13.078 & 0.01 & \(-\)1.6 & \(-\)3.948 & 0.030\\
NGC 2298* & 102.248 & \(-\)36.008 & 9.828 & 0.14 & \(-\)1.92 & \(-\)3.255 & 0.038\\
NGC 2808 & 138.084 & \(-\)64.872 & 10.06 & 0.22 & \(-\)1.14 & \(-\)4.239 & 0.029\\
NGC 3201 & 154.746 & \(-\)46.577 & 4.737 & 0.24 & \(-\)1.59 & \(-\)4.046 & 0.023\\
NGC 4590 & 189.852 & \(-\)26.725 & 10.404 & 0.05 & \(-\)2.23 & \(-\)3.939 & 0.021\\
NGC 4833 & 194.771 & \(-\)70.914 & 6.48 & 0.32 & \(-\)1.85 & \(-\)4.010 & 0.028\\
NGC 5024 & 198.189 & 18.191 & 18.498 & 0.02 & \(-\)2.1 & \(-\)4.085 & 0.022\\
NGC 5053* & 199.051 & 17.772 & 17.537 & 0.01 & \(-\)2.27 & \(-\)3.399 & 0.029\\
NGC 5139 & 201.699 & \(-\)47.491 & 5.426 & 0.12 & \(-\)1.53 & \(-\)4.028 & 0.019\\
NGC 5272 & 205.528 & 28.349 & 10.175 & 0.01 & \(-\)1.5 & \(-\)4.052 & 0.020\\
NGC 5466 & 211.557 & 28.503 & 16.12 & 0 & \(-\)1.98 & \(-\)4.008 & 0.025\\
NGC 5897 & 229.375 & \(-\)21.035 & 12.549 & 0.09 & \(-\)1.9 & \(-\)3.934 & 0.051\\
NGC 5904 & 229.650 & 2.110 & 7.479 & 0.03 & \(-\)1.29 & \(-\)4.127 & 0.020\\
NGC 5927 & 232.003 & \(-\)50.785 & 8.27 & 0.45 & \(-\)0.49 & \(-\)3.495 & 0.035\\
NGC 5986 & 236.483 & \(-\)37.783 & 10.54 & 0.28 & \(-\)1.59 & \(-\)4.204 & 0.031\\
NGC 6093 & 244.273 & \(-\)22.973 & 10.339 & 0.18 & \(-\)1.75 & \(-\)4.065 & 0.024\\
NGC 6101 & 246.564 & \(-\)72.187 & 14.449 & 0.05 & \(-\)1.98 & \(-\)3.899 & 0.028\\
NGC 6121 & 245.860 & \(-\)26.789 & 1.851 & 0.44 & \(-\)1.16 & \(-\)3.472 & 0.035\\
NGC 6144 & 246.803 & \(-\)26.026 & 8.151 & 0.63 & \(-\)1.76 & \(-\)4.117 & 0.034\\
NGC 6171 & 248.130 & \(-\)13.049 & 5.631 & 0.33 & \(-\)1.02 & \(-\)3.658 & 0.029\\
NGC 6205 & 250.424 & 36.447 & 7.419 & 0.02 & \(-\)1.53 & \(-\)4.027 & 0.024\\
NGC 6218 & 251.800 & \(-\)1.922 & 5.109 & 0.19 & \(-\)1.37 & \(-\)3.900 & 0.021\\
NGC 6254 & 254.312 & \(-\)4.097 & 5.067 & 0.28 & \(-\)1.56 & \(-\)4.063 & 0.027\\
NGC 6341 & 259.340 & 43.215 & 8.501 & 0.02 & \(-\)2.31 & \(-\)3.985 & 0.032\\
NGC 6352 & 261.505 & \(-\)48.439 & 5.543 & 0.22 & \(-\)0.64 & \(-\)3.280 & 0.066\\
NGC 6362 & 262.963 & \(-\)67.045 & 7.649 & 0.09 & \(-\)0.99 & \(-\)3.740 & 0.028\\
NGC 6366 & 261.893 & \(-\)5.155 & 3.444 & 0.71 & \(-\)0.82 & \(-\)3.210 & 0.034\\
NGC 6397 & 265.084 & \(-\)53.700 & 2.482 & 0.18 & \(-\)2.02 & \(-\)3.848 & 0.017\\
NGC 6541 & 272.089 & \(-\)43.678 & 7.609 & 0.14 & \(-\)1.81 & \(-\)4.063 & 0.049\\
NGC 6656 & 279.034 & \(-\)23.917 & 3.303 & 0.34 & \(-\)1.7 & \(-\)4.074 & 0.028\\
NGC 6723 & 284.923 & \(-\)36.575 & 8.267 & 0.05 & \(-\)1.1 & \(-\)3.851 & 0.040\\
NGC 6752 & 287.739 & \(-\)59.968 & 4.125 & 0.04 & \(-\)1.54 & \(-\)3.929 & 0.024\\
NGC 6779 & 289.152 & 30.146 & 10.43 & 0.26 & \(-\)1.98 & \(-\)3.817 & 0.030\\
NGC 6809 & 295.017 & -30.972 & 5.348 & 0.08 & \(-\)1.94 & \(-\)3.935 & 0.022\\
NGC 6838 & 298.483 & 18.787 & 4.001 & 0.25 & \(-\)0.78 & \(-\)3.573 & 0.064\\
NGC 7078 & 322.509 & 12.189 & 10.709 & 0.1 & \(-\)2.37 & \(-\)4.087 & 0.020\\
NGC 7089 & 323.372 & \(-\)0.799 & 11.693 & 0.06 & \(-\)1.65 & \(-\)3.964 & 0.021\\
NGC 7099 & 325.089 & \(-\)23.164 & 8.458 & 0.03 & \(-\)2.27 & \(-\)4.186 & 0.024\\
Arp 2 & 292.183 & \(-\)30.359 & 28.726 & 0.1 & \(-\)1.75 & \(-\)3.475 & 0.026\\
IC 4499 & 224.512 & \(-\)82.196 & 18.891 & 0.23 & \(-\)1.53 & \(-\)4.215 & 0.030\\
Lynga 7 & 242.768 & \(-\)55.315 & 7.899 & 0.73 & \(-\)0.67 & \(-\)3.870 & 0.044\\
NGC 2419 & 114.548 & 38.866 & 88.471 & 0.08 & \(-\)2.15 & \(-\)4.200 & 0.069\\
NGC 4372 & 186.642 & \(-\)72.619 & 5.713 & 0.39 & \(-\)2.17 & \(-\)3.925 & 0.080\\
NGC 5286 & 206.620 & \(-\)51.381 & 11.096 & 0.24 & \(-\)1.69 & \(-\)3.987 & 0.029\\
NGC 5634 & 217.408 & \(-\)5.980 & 25.959 & 0.05 & \(-\)1.88 & \(-\)4.114 & 0.052\\
NGC 5694 & 219.897 & \(-\)26.532 & 34.84 & 0.09 & \(-\)1.98 & \(-\)3.615 & 0.047\\
NGC 5824 & 225.972 & \(-\)33.041 & 31.713 & 0.13 & \(-\)1.91 & \(-\)3.905 & 0.041\\
NGC 5946 & 233.865 & \(-\)50.656 & 9.642 & 0.54 & \(-\)1.29 & \(-\)3.901 & 0.116\\
NGC 6139 & 246.924 & \(-\)38.842 & 10.035 & 0.75 & \(-\)1.65 & \(-\)3.836 & 0.102\\
NGC 6229 & 251.725 & 47.560 & 30.106 & 0.01 & \(-\)1.47 & \(-\)3.860 & 0.213\\
NGC 6235 & 253.345 & \(-\)22.136 & 11.937 & 0.31 & \(-\)1.28 & \(-\)3.815 & 0.075\\
NGC 6284 & 256.107 & \(-\)24.770 & 14.208 & 0.28 & \(-\)1.26 & \(-\)3.993 & 0.071\\
NGC 6304 & 258.638 & \(-\)29.430 & 6.152 & 0.54 & \(-\)0.45 & \(-\)3.301 & 0.052\\
NGC 6316 & 259.138 & \(-\)28.133 & 11.152 & 0.54 & \(-\)0.45 & \(-\)3.457 & 0.081\\
NGC 6325 & 259.495 & \(-\)23.782 & 7.533 & 0.91 & \(-\)1.25 & \(-\)3.753 & 0.097\\
NGC 6333 & 259.740 & \(-\)18.572 & 8.3 & 0.38 & \(-\)1.77 & \(-\)3.784 & 0.048\\
NGC 6342 & 260.306 & \(-\)19.572 & 8.013 & 0.46 & \(-\)0.55 & \(-\)3.678 & 0.069\\
NGC 6356 & 260.916 & \(-\)17.812 & 15.656 & 0.28 & \(-\)0.4 & \(-\)3.783 & 0.129\\
NGC 6388 & 264.183 & \(-\)44.787 & 11.171 & 0.37 & \(-\)0.75 & \(-\)3.316 & 0.086\\
NGC 6402 & 264.411 & \(-\)3.267 & 9.144 & 0.6 & \(-\)0.55 & \(-\)4.075 & 0.130\\
NGC 6453 & 267.715 & \(-\)34.610 & 10.07 & 0.64 & \(-\)0.46 & \(-\)3.855 & 0.056\\
NGC 6496 & 269.733 & \(-\)44.239 & 9.641 & 0.15 & \(-\)1.5 & \(-\)3.450 & 0.034\\
NGC 6522 & 270.871 & \(-\)30.047 & 7.295 & 0.48 & \(-\)0.46 & \(-\)3.525 & 0.096\\
NGC 6535 & 270.955 & \(-\)0.285 & 6.363 & 0.34 & \(-\)1.34 & \(-\)3.873 & 0.042\\
NGC 6539 & 271.212 & \(-\)7.571 & 8.165 & 1.02 & \(-\)1.79 & \(-\)3.765 & 0.107\\
NGC 6558 & 272.567 & \(-\)31.750 & 7.474 & 0.44 & \(-\)0.18 & \(-\)3.829 & 0.082\\
NGC 6569\textsuperscript{\textdagger *} & 273.439 & \(-\)31.838 & 10.534 & 0.53 & \(-\)0.76 & \(-\)3.806 & 0.056\\
NGC 6584 & 274.648 & \(-\)52.205 & 13.611 & 0.1 & \(-\)1.5 & \(-\)3.969 & 0.028\\
NGC 6624 & 275.943 & \(-\)30.327 & 8.019 & 0.28 & \(-\)0.44 & \(-\)3.540 & 0.029\\
NGC 6626 & 276.165 & \(-\)24.891 & 5.368 & 0.4 & \(-\)1.32 & \(-\)4.136 & 0.045\\
NGC 6637 & 277.840 & \(-\)32.331 & 8.9 & 0.18 & \(-\)0.64 & \(-\)3.952 & 0.026\\
NGC 6638 & 277.741 & \(-\)25.497 & 9.775 & 0.41 & \(-\)0.95 & \(-\)4.019 & 0.081\\
NGC 6652 & 278.924 & \(-\)32.986 & 9.464 & 0.09 & \(-\)0.81 & \(-\)3.753 & 0.032\\
NGC 6681 & 280.816 & \(-\)32.283 & 9.362 & 0.07 & \(-\)1.62 & \(-\)4.012 & 0.025\\
NGC 6712 & 283.337 & \(-\)8.713 & 7.382 & 0.45 & \(-\)1.02 & \(-\)4.016 & 0.113\\
NGC 6717 & 283.797 & \(-\)22.674 & 7.524 & 0.22 & \(-\)1.26 & \(-\)4.212 & 0.038\\
NGC 6760 & 287.809 & 1.039 & 8.411 & 0.77 & \(-\)0.4 & \(-\)3.612 & 0.112\\
NGC 6864 & 301.500 & \(-\)21.920 & 20.517 & 0.16 & \(-\)1.29 & \(-\)4.156 & 0.047\\
NGC 6934 & 308.550 & 7.390 & 15.716 & 0.1 & \(-\)1.47 & \(-\)4.066 & 0.028\\
NGC 6981 & 313.370 & \(-\)12.538 & 16.661 & 0.05 & \(-\)1.42 & \(-\)4.035 & 0.024\\
NGC 7006 & 315.372 & 16.180 & 39.318 & 0.05 & \(-\)1.52 & \(-\)3.977 & 0.031\\
NGC 7492 & 347.093 & \(-\)15.628 & 24.39 & 0 & \(-\)1.78 & \(-\)3.595 & 0.051\\
Pal 4 & 172.311 & 28.971 & 101.39 & 0.01 & \(-\)1.41 & \(-\)3.630 & 0.055\\
Pal 8 & 280.385 & \(-\)19.839 & 11.316 & 0.32 & \(-\)0.37 & \(-\)3.203 & 0.122\\
Pal 12 & 326.647 & \(-\)21.239 & 18.494 & 0.02 & \(-\)0.85 & \(-\)3.426 & 0.035\\
Rup 106 & 189.652 & \(-\)51.140 & 20.711 & 0.2 & \(-\)1.68 & \(-\)3.785 & 0.048\\
Terzan 3 & 247.094 & \(-\)35.363 & 7.644 & 0.73 & \(-\)0.74 & \(-\)3.682 & 0.090\\
Terzan 7 & 289.413 & \(-\)34.651 & 24.278 & 0.07 & \(-\)0.32 & \(-\)3.639 & 0.044\\
Terzan 8 & 295.500 & \(-\)33.972 & 27.535 & 0.12 & \(-\)2.16 & \(-\)3.854 & 0.033\\
\hline
\multicolumn{8}{l}{* denotes the TRGB not selected for the following analysis and }\\
\multicolumn{8}{l}{\textdagger\ denotes a possible AGB }\\
\end{longtable}
\subsection{Discussion}
\label{sec:orgda32828}

\subsubsection{Relationship between TRGB absolute magnitude and metallicity}
\label{sec:orgacfb50e}

\begin{figure}[!t]
\centering
\includegraphics[width=.9\linewidth]{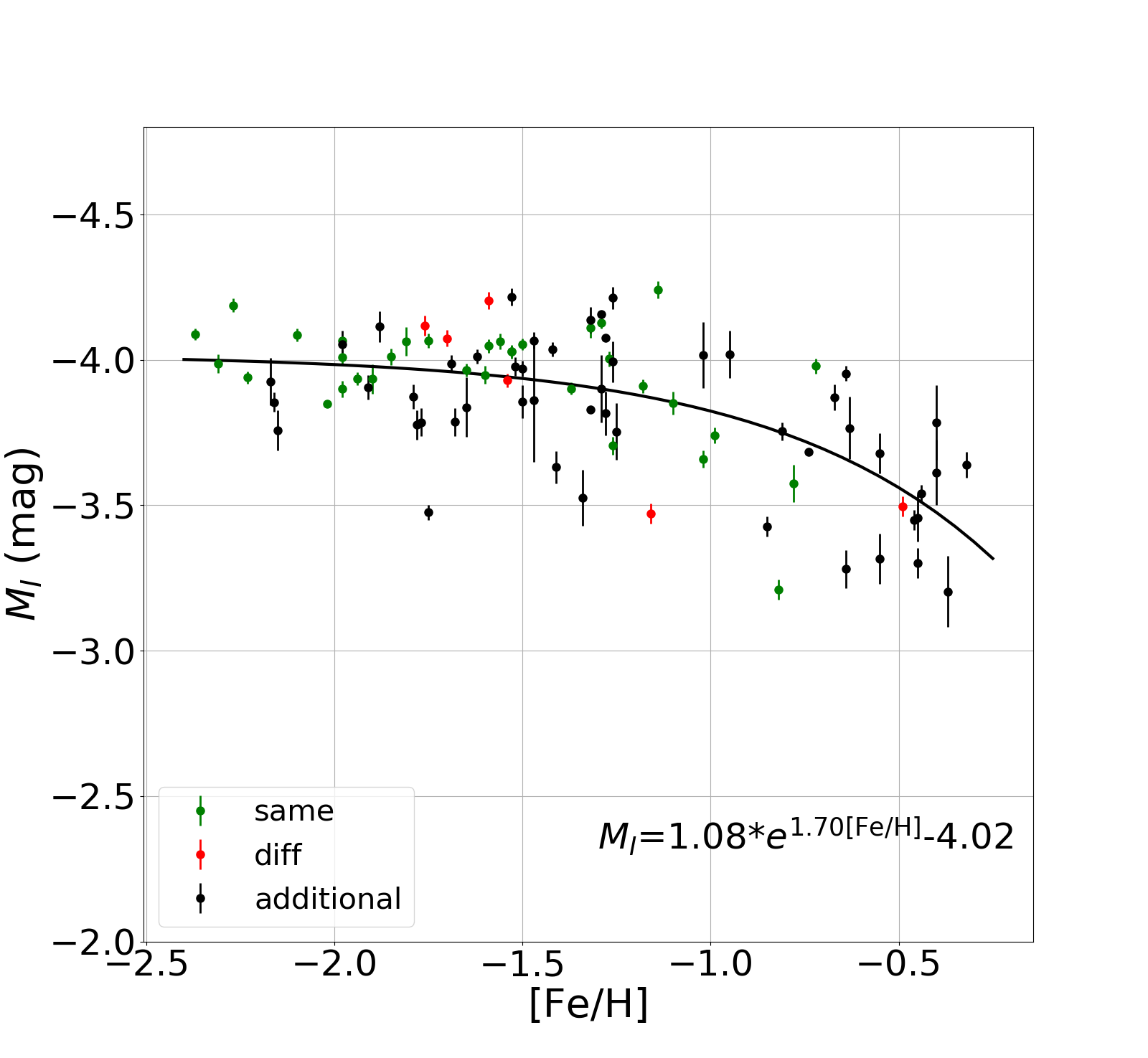}
\caption{\label{fig:result-mag-meta}
Absolute magnitudes vs metallicity of TRGBs, showing TRGB samples identified consistently by both machine recognition and manual selection (green dots), inconsistent samples (red dots), and newly added star clusters (black triangles).}
\end{figure}

The absolute magnitude in the I band of the TRGB is a relatively
stable constant, often used as a standard candle for distance
measurement. However, stellar evolution theory reveals that the I band
absolute magnitude depends on metallicity
\citep{2017-Serenelli-Brightness-A}. At low metallicities, the I-band
magnitude of the TRGB remains nearly constant. However, as metallicity
increases, the I-band magnitude gradually fades. To quantify the
effect of metallicity on the TRGB's absolute magnitude, we plotted the
relationship between TRGB metallicity and absolute magnitude using the
previously calculated results. As shown in Figure \ref{fig:result-mag-meta},
the analysis reveals a clear trend in I-band luminosity depending on
metallicity.

For lower metallicities ([Fe/H] < \(-\)1), I-band luminosity remains
relatively constant. At higher metallicities ([Fe/H] > \(-\)1), we
observe a systematic I-band luminosity decrease. Our fitting yields an
asymptotic constant \(-\)4.02 for absolute magnitude of I band in
metal-poor region, consistent with previous studies
\citep{2021-Freedman-Measurements-A,2019-Freedman-CarnegieChicago-A,2021-Hoyt-Carnegie-A,2019-Yuan-Consistent-A,2021-Anand-Distances-M}.
This consistency validates our algorithm's effectiveness, with most of
the automated selected TRGB samples matching manual selections from
\cite{2025-Shao-Dependence-A}. We excluded three samples (the globular
clusters marked with an asterisk in Table \ref{tab:cluster_trgb}) from the
initial set of 94 TRGB sources: two beyond 3\(\sigma\) of our fit and one
potential AGB star. The two deviant sources, also flagged by
\cite{2025-Shao-Dependence-A}, with a deviation from the trend line
locating at the upper extremity, suggesting that these stars are
generally warmer and may not have reached the TRGB stage in their
evolution \citep{2025-Shao-Dependence-A}.

Previous investigations of TRGB metallicity dependence focused on
relation between bolometric luminosity and [Fe/H] using photographic
photometry of about 6-8 Galactic globular clusters by selected the
brightest RGB source as TRGB
\citep{1990-DaCosta-Standard-A,2000-Ferraro-New-A}. Building on this,
\cite{2001-Bellazzini-Step-A} derived an M\textsubscript{I}-[Fe/H] relation by
combining bolometric magnitude correlations from
\cite{2000-Ferraro-New-A} with color index and bolometric correction
relations from \cite{1990-DaCosta-Standard-A}. While these studies
focused on bolometric luminosity correlations, their M\textsubscript{I}-[Fe/H]
relationship remains a semi-empirical derivation. Our analysis
benefits from Gaia DR3's \citep{2023-GaiaCollaboration-Gaia-A}
complete globular cluster photometry, covering wider metallicity
ranges than previous studies. With 91 clusters analyzed - almsot an
order of magnitude improvement over typical samples of <10 clusters -
we achieve richer metallicity coverage while maintaining astrometric
precision for reliable TRGB detection.

\begin{figure}[!t]
\centering
\includegraphics[width=.9\linewidth]{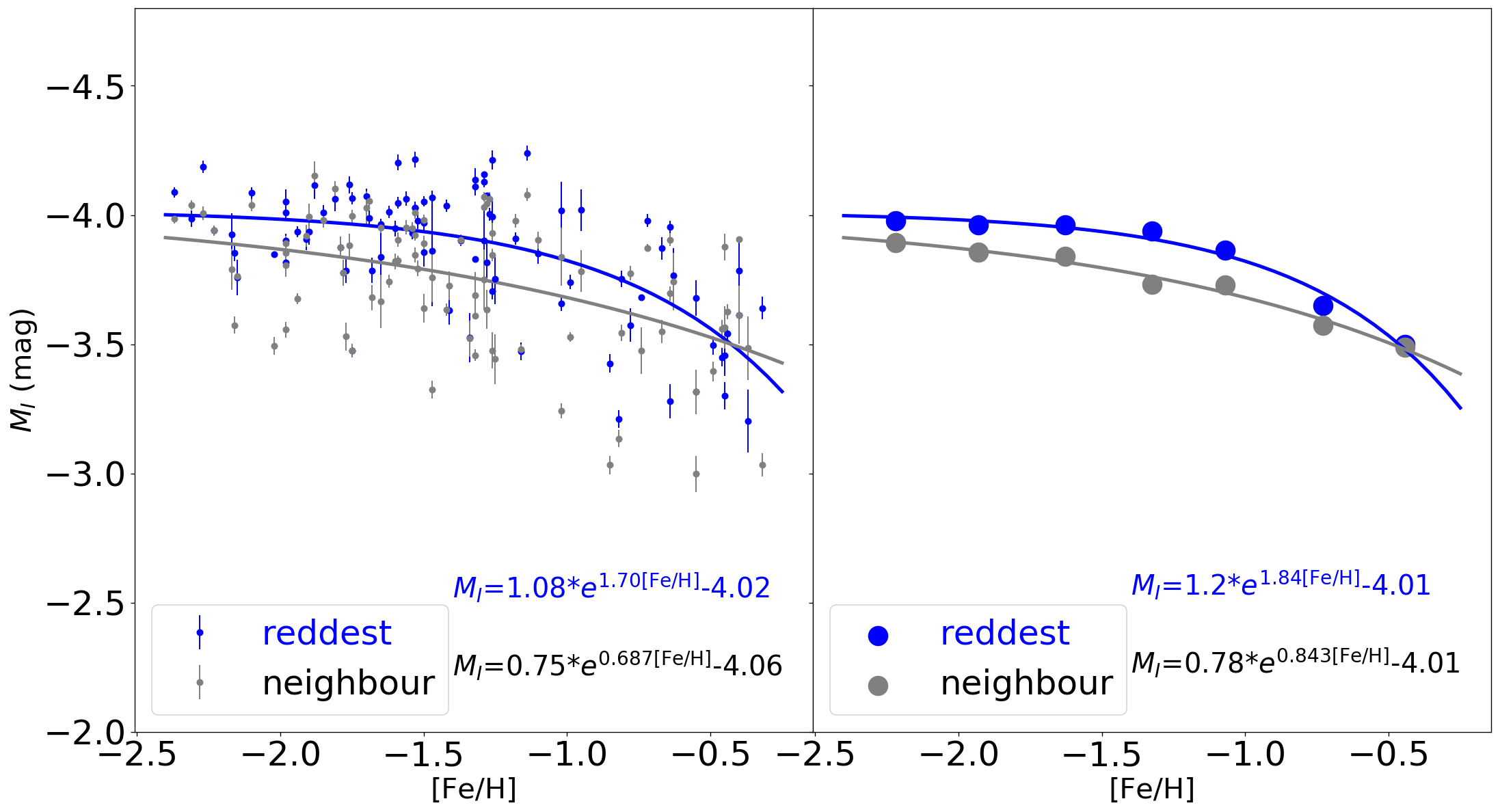}
\caption{\label{fig:result-two-mag-meta-bin}
Absolute magnitudes vs metallicity of TRGBs. Left panel: The blue dots represent data from selected samples, gray dots represent closest neighbors from each cluster's selected sample, blue and gray lines denote the fitting curves repectively; Right panel: Same as left panel with M\textsubscript{I} averaged in [Fe/H] bins of 0.3 dex intervals.}
\end{figure}

To evaluate the systematic error of the method on TRGB absolute
magnitude, we used its neighbour (the nearest star to each selected
sample) in the CMD of each globular cluster to calculate its I-band
absolute magnitude and compared it to the selected sample. As shown in
Figure \ref{fig:result-two-mag-meta-bin}, we plotted the fitting curves for
both the full sample and the binned sample along with the mean
absolute magnitudes binned by [Fe/H] with a size of 0.3 dex. Although
the distribution between the reddest and their neighbour sources show
a gap, the TRGB limiting magnitudes fitted for all sources become very
close: \(-\)4.02 and \(-\)4.06 respectively, when metallicity
approaches infinitesimal values. When fitting by bin, the derived TRGB
limiting magnitudes are nearly identical at \(-\)4.01. By binning the
fitting, the influence of the single TRGB metering error can be
offset, and the fitting results are more consistent. This demonstrates
that our fitting results are quite robust and consistent.

The TRGB limiting magnitudes difference of 0.04 between the reddest
and their neighbour sources can be used as the systematic error of the
method on TRGB absolute magnitude. This indicates that the results
have good consistency, and the impact of error caused by the method is
small.

Comparing the reddest and their neighbour sources reveals that when
metallicity is low, the brightness of the reddest sources and the
neighbour are similar in most clusters, with the neighbour being
slightly dimmer which cause the neighbour fiting curve lower than the
reddest one. However, at higher metallicities, especially above
\(-\)0.6, the neighbour becomes brighter than the reddest source in
some clusters. This aligns with the effect mentioned previously: as
metallicity increases, there is a downward bending effect near the end
of the red giant branchs in CMDs, resulting in brighter neighbour
sources. The disparity between the reddest and the neighbour sources
primarily stems from the limited number of stars on the red giant
branch and photometric errors. In clusters with sparser red giant
branches, the brightness difference between the neighbour and the
reddest one is larger than that in the dense clusters.
\subsubsection{Relationship between TRGB color index and metallicity}
\label{sec:org82b87d2}

\begin{figure}[!t]
\centering
\includegraphics[width=.9\linewidth]{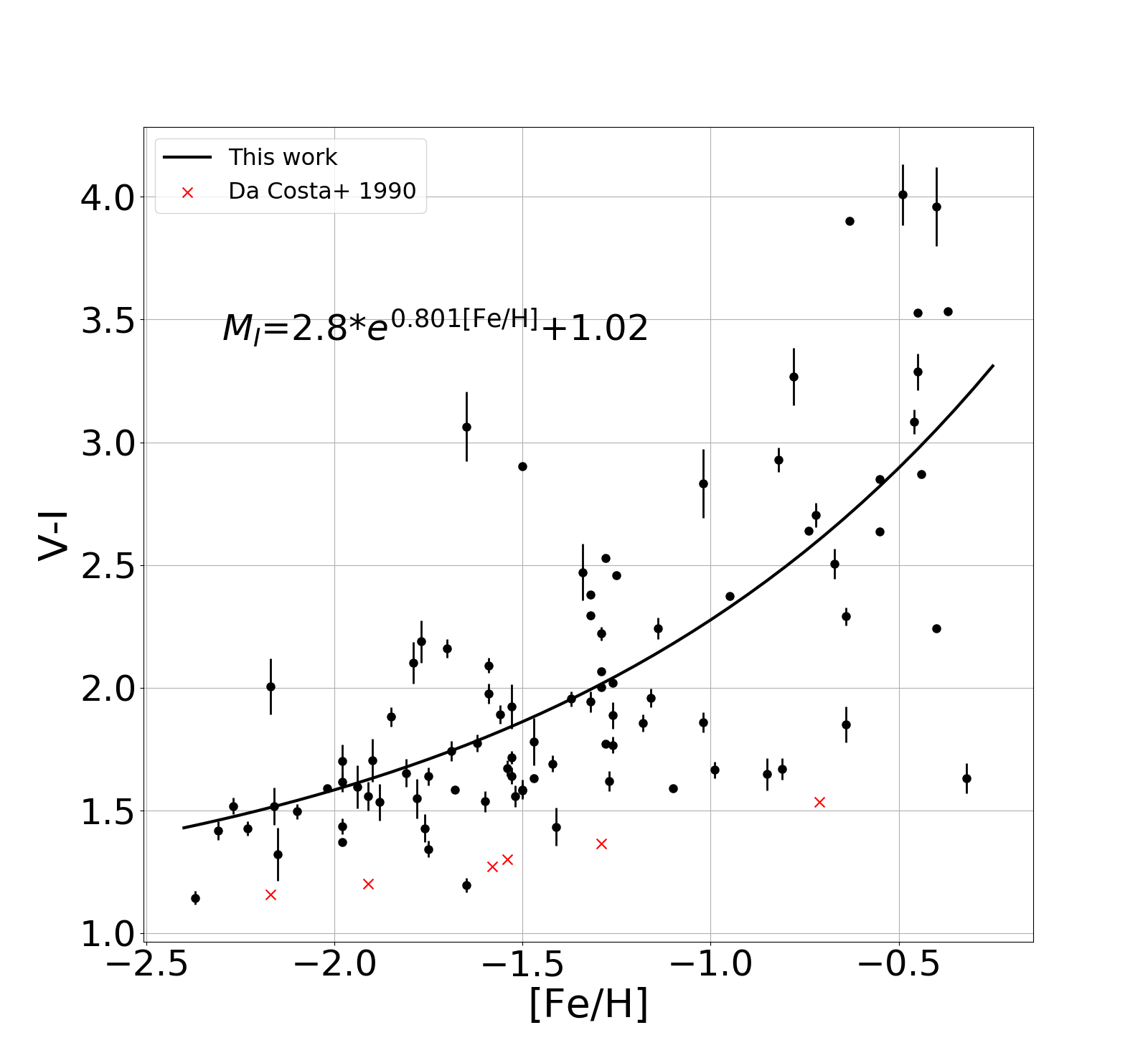}
\caption{\label{fig:result-BP-RP-meta}
Relationship between metallicity ([Fe/H]) and color index (V \(-\) I) of TRGB stars, with the fitted curve representing the trend. Red x markers denote samples in \cite{1990-DaCosta-Standard-A}, and black dots denote our samples.}
\end{figure}

As shown in Figure \ref{fig:result-BP-RP-meta}, by analyzing the TRGB
results, we found a trend between metallicity ([Fe/H]) and color (V
\(-\) I). The data suggests that as metallicity increases, the V \(-\) I
color index of the TRGB tends to increase. This indicates that
metal-rich TRGB stars are generally redder than their metal-poor
counterparts.  This trend is consistent with stellar evolution theory
that the extra metals in the atmospheres of metal-rich stars cause
opacity in the convective atmosphere, making the star trap heat,
expand and become cooler \citep{2012-Bressan-Parsec-M}. This is
consistent with the trends marked as red x in
\cite{1990-DaCosta-Standard-A}. While the data in
\cite{1990-DaCosta-Standard-A} is selected with M\textsubscript{I} near \(-\)3 and
our data is near \(-\)4, there is a upward shift as our samples is
redder. The relationship between metallicity and color can be used to
refine TRGB calibration and further distance measurements.
\subsubsection{Performance and Applications of the Automated Method}
\label{sec:org3881430}

This study successfully achieved the automatic identification and
analysis of TRGB in globular clusters through the combination of
automated scripts and algorithms. Compared to traditional manual
methods, the automated approach not only improves efficiency but also
aims to reduce the impact of artificial selection. Notably, our
automated method expands the TRGB sample by over 100\% (91 vs. 37),
with a threefold increase in the metal-rich range (\(-\)1 < [Fe/H]
< 0) where metallicity effects are most significant
\citep{2025-Shao-Dependence-A}. The close agreement between our TRGB
absolute magnitude–metallicity relation and previous results further
validates the robustness and reliability of our approach.

Taking into account upcoming high-precision instruments such as JWST
and CSST, while traditional manual identification methods are
inefficient when dealing with the massive photometric datasets for
star clusters in nearby galaxies like the Andromeda Galaxy (M31), the
Large and Small Magellanic Clouds (LMC/SMC), the proposed automated
TRGB identification method can address challenges. Our automated
pipeline provides a scalable and objective framework capable of
processing large volumes of high-resolution data with minimal human
intervention. Therefore, this approach not only resolves the
inefficiency and subjectivity of previous methods but also lays the
foundation for automated, large-scale TRGB mapping in nearby galaxies.

However, this study also has some limitations. Firstly, the parameter
selection of the LOF algorithm significantly affects the effectiveness
of outlier identification, and further research is needed to optimize
the parameters. Secondly, the determination of the TRGB depends on the
accuracy of the fitted curve, and the choice of the fitting curve may
affect the final result. Therefore, in future research, we plan to
explore more advanced machine learning algorithms to improve the
accuracy of TRGB identification. Additionally, expanding the sample
size and exploring other data sources could further enhance the
robustness and generalizability of our findings.
\section{Conclusions}
\label{sec:orgdb6351d}

In conclusion, this study presents an automated methodology for
identifying the TRGB in globular clusters. By integrating astronomical
data with algorithmic techniques including data filtering and
cross-matching, CMD construction, outlier removal, and second-degree
polynomial fitting, we have developed an efficient and accurate
process for TRGB identification. Finally, 91 TRGBs were identified
from 160 global clusters, which extends samples in previous studies,
especillay on the high metallicity region. The results indicate a
robust \(\rm M_{I}\) fit around \(-\)4.02 at extremely low
metallicities, with the I-band luminosity decreasing for metallicities
above \(-\)1. Our automated approach substantially enhances the
efficiency of TRGB identification, enabling its effective application
to large-scale datasets and forthcoming surveys such as CSST.
\section*{Acknowledgments}
\label{sec:orgcf143f6}
This work is supported by the National Natural Science Foundation of
China (No. 32300381, No. 12503027, No. 12303035, No. 12203007), Young
Data Scientist Program of the China National Astronomical Data Center
(No. NADC2025YDS-02), Innovation Ideas of Beijing Planetarium, Popular
Science Project of Beijing Planetarium, China Manned Space Program
with grant No. CMS-CSST-2025-A01, Beijing Natural Science Foundation
(No. 1242016), and Talents Program (24CE-YS-08), Popular Science
Project (24CD012), Innovation Project (24CD013), Mengya Program
(BGS202203) of Beijing Academy of Science and Technology.

\bibliographystyle{raa}
\bibliography{ms2025-0314.bib}
\end{document}